%% file: wrapper.tex
\documentclass[a4paper,11pt]{report}
\usepackage[utf8]{inputenc}
\usepackage[T1]{fontenc}
\usepackage{amsmath,amssymb,array}
\usepackage{booktabs}
\usepackage{subcaption}
\usepackage{graphicx}

\makeatletter

\RequirePackage{geometry}
\RequirePackage{fancyhdr}
\fancyheadoffset{2cm} 
\RequirePackage{microtype}

\RequirePackage[scaled=0.92]{helvet}
\RequirePackage{palatino,mathpazo}
\RequirePackage[scaled=1.02]{inconsolata}
\RequirePackage[T1]{fontenc}

\RequirePackage[hyphens]{url}
\RequirePackage[pagebackref]{hyperref}

\RequirePackage{color}
\definecolor{link}{rgb}{0.45,0.51,0.67}
\hypersetup{
  colorlinks,%
  citecolor=link,%
  filecolor=link,%
  linkcolor=link,%
  urlcolor=link
}

\RequirePackage{setspace}
\newcommand{\volume}[1]{\def\RJ@volume{#1}}
\newcommand{\volnumber}[1]{\def\RJ@number{#1}}
\renewcommand{\month}[1]{\def\RJ@month{#1}}
\renewcommand{\year}[1]{\def\RJ@year{#1}}

\newcommand  {\sectionhead}  [1]{\def\RJ@sectionhead{#1}}
\renewcommand{\author}       [1]{\def\RJ@author{#1}}
\renewcommand{\title}        [1]{\def\RJ@title{#1}}
\newcommand  {\subtitle}     [1]{\def\RJ@subtitle{#1}}

\usepackage[medium]{titlesec}
\usepackage{titletoc}
\renewcommand{\thesection}{\arabic{section}}
\renewcommand{\thesubsection}{\arabic{section}.\arabic{subsection}}
\titleformat{\section}{\normalfont\large\bfseries}{\thesection}{1em}{}
\titleformat{\subsection}{\normalfont\normalsize\bfseries}{\thesubsection}{1em}{}
\titlecontents{chapter}   [0em]{}{}{}{\titlerule*[1em]{.}\contentspage}

\RequirePackage{placeins}
\newenvironment{article}{\author{}\title{}\subtitle{}\FloatBarrier}{\FloatBarrier}

\renewcommand{\abstract}[1]{%
\setstretch{1}%
\noindent%
\small%
\textbf{Abstract} #1 
}

\renewcommand{\maketitle}{%
\noindent
  \chapter{\RJ@title}\refstepcounter{chapter}
  \ifx\empty\RJ@subtitle
  \else
    \noindent\textbf{\RJ@subtitle}
    \par\nobreak\addvspace{\baselineskip}
  \fi
  \ifx\empty\RJ@author
  \else
    \noindent\textit{\RJ@author}
    \par\nobreak\addvspace{\baselineskip}
  \fi
  \@afterindentfalse\@nobreaktrue\@afterheading
}

\renewcommand\chapter{\secdef\RJ@chapter\@schapter}

\newcommand{\RJ@chapter}{%
  \edef\name@of@eq{equation.\@arabic{\c@chapter}}%
  \renewcommand{\@seccntformat}[1]{}%
  \@startsection{chapter}{0}{0mm}{%
    -2\baselineskip \@plus -\baselineskip \@minus -.2ex}{\p@}{%
    \phantomsection\normalfont\huge\bfseries\raggedright}}

\RequirePackage[sectionbib,round]{natbib}
\renewcommand{\theequation}{\@arabic\c@equation}
\renewcommand{\thefigure}{\@arabic\c@figure}
\renewcommand{\thetable}{\@arabic\c@table}

\renewcommand{\contentsname}{Contents}
\renewcommand\tableofcontents{%
   \vspace{1cm}
   \section*{\contentsname}
   { \@starttoc{toc} }
}

\renewcommand{\titlepage}{%
  \thispagestyle{empty}
  \hypersetup{
    pdftitle={The R Journal Volume \RJ@volume/\RJ@number, \RJ@month \RJ@year},%
    pdfauthor={R Foundation for Statistical Computing},%
  }
  \noindent
  \begin{center}
    \fontsize{50pt}{50pt}\selectfont
    The \raisebox{-8pt}{\includegraphics[height=77pt]{Rlogo-5}}\hspace{10pt}
    Journal
  
  \end{center}
  {\large \hfill Volume \RJ@volume/\RJ@number, \RJ@month{} \RJ@year \quad}

  \rule{\textwidth}{1pt}
  \begin{center}
    {\Large A peer-reviewed, open-access publication of the \\
    R Foundation for Statistical Computing}
  \end{center}

  \setcounter{tocdepth}{0}
  \tableofcontents
  \setcounter{tocdepth}{2}
  \clearpage
}

\newcommand{\address}[1]{\addvspace{\baselineskip}\noindent\emph{#1}}
\newcommand{\email}[1]{\href{mailto:#1}{\normalfont\texttt{#1}}}

\DeclareRobustCommand\code{\bgroup\@noligs\@codex}
\def\@codex#1{\texorpdfstring%
{{\normalfont\ttfamily\hyphenchar\font=-1 #1}}%
{#1}\egroup}

\DeclareRobustCommand\samp{`\bgroup\@noligs\@sampx}
\def\@sampx#1{{\normalfont\texttt{#1}}\egroup'}

\newcommand{\dfn}[1]{{\normalfont\textsl{#1}}}

\let\pkg=\strong
\newcommand{\CRANpkg}[1]{\href{https://CRAN.R-project.org/package=#1}{\pkg{#1}}}%

\RequirePackage{fancyvrb}
\RequirePackage{alltt}

\DefineVerbatimEnvironment{example}{Verbatim}{}
\renewenvironment{example*}{\begin{alltt}}{\end{alltt}}

\DefineVerbatimEnvironment{Sinput}{Verbatim}{}
\DefineVerbatimEnvironment{Soutput}{Verbatim}{}
\DefineVerbatimEnvironment{Scode}{Verbatim}{}
\DefineVerbatimEnvironment{Sin}{Verbatim}{}
\DefineVerbatimEnvironment{Sout}{Verbatim}{}

\providecommand{\operatorname}[1]{%
  \mathop{\operator@font#1}\nolimits}
\RequirePackage{amsfonts}

\renewcommand{\P}{%
  \mathop{\operator@font I\hspace{-1.5pt}P\hspace{.13pt}}}
\newcommand{\E}{%
  \mathop{\operator@font I\hspace{-1.5pt}E\hspace{.13pt}}}

\RequirePackage[font=small,labelfont=bf]{caption}

\begin{document}

\sectionhead{Dalitz, Lögler: moonboot}
\volume{}
\volnumber{}
\year{}
\month{}

\begin{article}
  \input{moonboot}

\end{article}

\end{document}

%% file: moonboot.tex
\renewcommand{\topfraction}{.8}

\title{moonboot: An R Package Implementing m-out-of-n Bootstrap Methods}
\author{by Christoph Dalitz and Felix L\"ogler}

\maketitle

\abstract{
The m-out-of-n bootstrap is a possible workaround to compute confidence intervals for bootstrap inconsistent estimators, because it works under weaker conditions than the n-out-of-n bootstrap. It has the disadvantage, however, that it requires knowledge of an appropriate scaling factor $\tau_n$ and that the coverage probability for finite $n$ depends on the choice of $m$. This article presents an R package \CRANpkg{moonboot} which implements the computation of m-out-of-n bootstrap confidence intervals and provides functions for estimating the parameters $\tau_n$ and $m$. By means of Monte Carlo simulations, we evaluate the different methods and compare them for different estimators.
}

\section{Introduction}
The \dfn{bootstrap} is a resampling method introduced by \citet{efron:1979}, which repeatedly simulates samples of the same size $n$ as the observed data by $n$-fold drawing with replacement. Let us call this method \dfn{n-out-of-n bootstrap}, in order to distinguish it from different sampling schemes. For each of these samples, the estimator $T$ of interest is computed, thereby simulating a sample $T_1^*, \ldots, T_R^*$ for the distribution of $T$, where $R$ is the number of bootstrap repetitions. For sufficiently smooth estimators $T$, this bootstrap distribution asymptotically approaches the true distribution of $T$ \citep{gine:1990,shao:1995}, and it can thus be used to construct confidence intervals for $T$. The coverage probability of confidence intervals based on flipping the quantiles of the bootstrapped distribution $T_1^*, \ldots, T_R^*$ at the point estimate (``basic bootstrap'') can be shown to be first order accurate, i.e.~up to $O(n^{-1/2})$. Under certain conditions, it is even possible to make the confidence intervals second order accurate, i.e.~up to $O(n^{-1})$, by studentized sampling (``bootstrap-$t$'') or accelerated bias correction (``$BC_a$ bootstrap'')  \citep{hall:1988,diciccio:1996,davison:1997}. Compared to other methods for estimating confidence intervals, the bootstrap is more versatile because it neither relies on a likelihood function, nor does it require asymptotic normality. The bootstrap has thus become a standard technique for estimating confidence intervals, and it is even included in vanilla R with the package \pkg{boot} \citep{R:boot}, which implements five different methods in \code{boot.ci()}. 

There are, however, \dfn{bootstrap inconsistent} estimators, i.e., estimators for which the n-out-of-n bootstrap fails to yield confidence intervals with an asymptotically correct coverage probability. Examples include extreme order statistics \citep{bickel:1997}, the Grenander estimator of a monotonous density \citep{sen:2010}, Chernoff's estimator of the mode and Tukey's shorth \citep{leger:2006}, or Chatterjee's rank correlation index \citep{lin:2023,dalitz:2024}. See the discussion by \citet{lin:2023} for a literature review and further examples. A possible workaround in these cases is the \dfn{m-out-of-n bootstrap}, which samples only $m<n$ observations. This can be done with or without replacement. \citet{politis:1994}, who introduced this method, only considered sampling without replacement, which they called \dfn{subsampling}. They have shown that the m-out-of-n bootstrap without replacement works under much weaker conditions than the n-out-of-n bootstrap, provided $m$ is chosen such that, for $n\to\infty$, $m\to\infty$ and $m/n\to 0$. The only condition is that the estimator, suitably scaled by some factor $\tau_n$, possesses a limit distribution, whereas no smoothness of the estimator or uniformity of the convergence is required. If the sampling is instead done with replacement, \cite{bickel:1997} have shown that a similar result requires additional restrictions on the estimator. Sampling without replacement thus works under weaker conditions than sampling with replacement, and, in the present article, we therefore only consider sampling without replacement. Our package \CRANpkg{moonboot}\footnote{This article used version 2.0.x, available, e.g., from \url{https://github.com/cdalitz/moonboot}.} supports sampling both with and without replacement by means of an argument \code{replace}, which is set to \code{FALSE} by default.

The wider applicability of the m-out-of-n bootstrap comes at a price, though. One drawback is that the scaling factor $\tau_n$ must be known. \citet{bertail:1999} explained that it must be chosen such that $\tau_n^2 \operatorname{Var}(T)$ converges to some constant, and they also suggested a method to estimate the scaling factor with another bootstrap. For root-$n$ consistent estimators, it is $\tau_n=\sqrt{n}$, but not all estimators are root-$n$ consistent, especially if they are bootstrap inconsistent, and even if root-$n$ consistency is conjectured it can be difficult to prove \citep{lin:2023}. Moreover, the m-out-of-n bootstrap can have less than first order accuracy: \citet{politis:1994} gave a theoretical example with accuracy $O(n^{-1/3})$ for an optimal choice of $m$, and even worse for other choices of $m$. That the accuracy depends on the choice of $m$ was demonstrated, too, in a simulation study by \citet{kleiner:2014}. And for the sample quantiles and sampling with replacement, \citet{arcones:2003} has shown that the choice $m\propto n^{2/3}$ is optimal. This raises the problem how to choose $m$, and several heuristics have been suggested in the literature \citep{politis:1999,goetze:2001,bickel:2008,chung:2001,lee:2020}.

Due to these shortcomings, the usual n-out-of-n bootstrap is generally preferable for bootstrap consistent estimators. However, there \emph{are} bootstrap inconsistent estimators and there is thus need for an R package that facilitates the application of the m-out-of-n bootstrap by providing the required functions. To this end, we present a new package \CRANpkg{moonboot}, evaluate the algorithms implemented therein by means of Monte Carlo simulations, and give recommendations and use cases for their application. The name of the package was inspired by an article by \citet{goetze:2001}, who abbreviated the m-out-of-n bootstrap as ``moon bootstrap''.

\section{The m-out-of-n bootstrap}
\label{sec:moon-bootstrap}
Let $X_1,\ldots,X_n$ be i.i.d.~random variables, $\theta$ be some parameter of their underlying distribution, and $T_n=T_n(X_1,\ldots,X_n)$ be an estimator for $\theta$. Let us additionally assume\footnote{\citet{politis:1994} did not make this assumption, but it is necessary for the estimation of the scaling factor $\tau_n$ with the method by \citet{bertail:1999}.} that $E(T_n^2)<\infty$. The m-out-of-n bootstrap requires that, for some scaling factor $\tau_n$ which behaves for $n\to\infty$ as
\begin{equation}
  \label{eq:tau}
  \frac{\tau_{m(n)}}{\tau_n} \to 0 \quad\mbox{ for } \quad m(n)\to\infty \quad\mbox{ with }\quad \frac{m(n)}{n}\to 0,
\end{equation}
the estimator converges in distribution to some limit law when centered around the true parameter $\theta$ and scaled with $\tau_n$:
\begin{equation}
  \label{eq:tscaled}
  S_n = \tau_n (T_n - \theta) \;\stackrel{n\to\infty}{\longrightarrow}\; S \;\mbox{ in distribution} 
\end{equation}
Note that condition (\ref{eq:tau}) implies that $\tau_n\to\infty$, and thus the convergence (\ref{eq:tscaled}) requires that $T_n$ converges to $\theta$ in probability (in other words: $T_n$ must be consistent), because it follows from (\ref{eq:tscaled}) that
$$P(|T_n-\theta|>\varepsilon) = F_n(-\tau_n\varepsilon)+(1-F_n(\tau_n\varepsilon)) \;\stackrel{n\to\infty}{\longrightarrow}\; 0$$
where $F_n$ is the cumulative distribution function of $S_n$. Moreover, if additionally the second moment of $S_n$ converges to some constant, the scaling factor $\tau_n$ is related to the rate of decrease of the variance of $T_n$:
\begin{equation}
  \label{eq:vartn}
  V = \lim_{n\to\infty} \operatorname{Var}(S_n) =
   \lim_{n\to\infty} \operatorname{Var}(\tau_n T_n)
\quad\Rightarrow\quad \operatorname{Var}(T_n) \sim \frac{V}{\tau_n^2} \quad\mbox{for}\quad n\to\infty
\end{equation}
This relationship allows to determine $\tau_n$ theoretically by means of an analytical calculation, or to estimate it by means of Monte Carlo simulations.

\citet{politis:1994} have shown that, under the assumptions (\ref{eq:tscaled}) and (\ref{eq:tau}), a confidence interval with asymptotic coverage probability $(1-\alpha)$ can be constructed as
\begin{equation}
  \label{eq:ci-m-out-of-n}
  \left[ T_n - \frac{q(1-\alpha/2)}{\tau_n},\; T_n - \frac{q(\alpha/2)}{\tau_n} \right]
\end{equation}
where $q(1-\alpha/2)$ and $q(\alpha/2)$ are the quantiles of the scaled bootstrap distribution $\tau_m(T_m^* - T_n)$, where $T_m^*$ denotes the bootstrap samples obtained by $m$-fold drawing without replacement. In \pkg{moonboot}, this interval is implemented by \code{mboot.ci(..., method="politis")}.

If the variance of $S_n$ converges to some value $\sigma^2$ and the limiting distribution of $S$ is the {\em normal distribution} with standard deviation $\sigma$, then the m-out-of-n bootstrap can alternatively be used to estimate the variance $\sigma^2$ and compute the standard confidence interval
\begin{equation}
  \label{eq:ci-norm}
  T_n \pm z_{1-\alpha/2}\,\hat{\sigma}_m^* \quad\mbox{with}\quad \hat{\sigma}_m^* = \frac{\tau_m}{\tau_n}\sqrt{\operatorname{Var}(T_m^*)}
\end{equation}
In \pkg{moonboot}, this is implemented by \code{mboot.ci(..., method="norm")}. Although this interval is only reasonable for asymptotically normal estimators, its coverage probability can have a faster convergence to the nominal value than the interval (\ref{eq:ci-m-out-of-n}) provided asymptotic normality indeed holds \citep{dalitz:2024}.

\citet{sherman:2004} observed that the scaling with $\tau_m/\tau_n$ in (\ref{eq:ci-m-out-of-n}) can be omitted if the interval is centered around one of the $T_m$ instead of $T_n$. They suggested to use the first $m$ samples for computing $T_m$, but this is an arbitrary choice and thus even the location of the confidence interval is not uniquely determined by the data. Moreover, the lack of scaling increases the interval length. The greater length is compensated by the higher volatility of the location so that the coverage probability still is close to the nominal value. This does not hold, however, for drawing without replacement as $m$ approaches $n$, and a compromise must be made between small interval length and approximate coverage probability. \citet{sherman:2004} suggested a heuristic method for choosing $m$ on basis of a double bootstrap, and they gave the rule of thumb to choose $m$ between $n^{1/2}$ and $n^{2/3}$. However, due to the greater interval length, this method should only be used as a last resort if the convergence rate $\tau_n$ is unknown and cannot be estimated from the data. In \pkg{moonboot}, this method is implemented by \code{mboot.ci(..., method="sherman")} and the method for estimating $m$ by \code{estimate.m.sherman()}.

\subsection{Estimation of $\tau_n$}
According to Eq.~(\ref{eq:vartn}), the formula for the scaling factor $\tau_n$ can be determined by an analytic investigation of $\operatorname{Var}(T_n)$, which can not only be difficult for some estimators, but it requires an ad hoc study of the specific estimator $T_n$ under consideration. This thwarts the application of the m-out-of-n bootstrap out-of-the box.

Fortunately, the relationship (\ref{eq:vartn}) makes it possible to estimate $\tau_n$ by another bootstrap \citep{bertail:1999}. If the variance $V_m=\operatorname{Var}(T_m^*)$ is estimated by sampling with different subsampling sizes $m$ and $\tau_n$ is assumed to be of the form $\tau_n=n^\beta$, then the asymptotic relationship (\ref{eq:vartn}) becomes
\begin{equation}
  \label{eq:bertail}
  \log V_m \approx -2\beta\log m + \log V
\end{equation}
and $\beta$ can be estimated with a least-squares fit. For the choice of the test values $m$, \citet{bertail:1999} suggested $m_i=n^{\gamma_i}$, but wrote that ``the difficult problem of choosing the $\gamma_i$'s requires more work on a case-by-case basis''\footnote{Note that \citet{bertail:1999} wrote $n^{\beta_i}$ for the test values $m_i$, but we have renamed them to $n^{\gamma_i}$ to avoid confusion with the power $\beta$ that is to be estimated.}. This casts doubt on the whole method, because removing the necessity of a detailed asymptotic analysis of the estimator is the whole point of estimating $\tau_n$ from the data. Indeed, we have not found a sequence of values for the $\gamma_i$ that worked equally well in all of our examples. We thus leave the choice of the $\gamma$-sequence as an option to the user and use the default of five values between $0.2$ and $0.7$, which worked reasonably well for some of our estimators.

\cite{bertail:1999} suggested the alternative of using quantiles instead of the variance, but as the method involves taking logarithms, this makes no sense when the estimator tends to be negative or close to zero. As a workaround, \citep{politis:1999} suggested to replace $V_m$ in Eq.~(\ref{eq:bertail}) with the average of quantile ranges
\begin{equation}
  Q_m = \frac{1}{J}\sum_{j=1}^J \left( \operatorname{quantile}(T_m^*, \alpha^{high}_j) - \operatorname{quantile}(T_m^*, \alpha^{low}_j) \right)
\end{equation}
We have implemented this, too, with the choice $\alpha^{high}_j=0.75+j\cdot 0.05$ and $\alpha^{low}_j=0.25-j\cdot 0.05$ for $j=0,\ldots,4$, but the results in section \ref{sec:results:tau} show that this makes almost no difference in comparison to using the variance.

This method is implemented in \pkg{moonboot} by \code{estimate.tau()}, which is automatically called by \code{mboot.ci()} when no rate for the parameter \code{tau} is provided.

\begin{figure}[t]
  \begin{subfigure}[c]{0.5\textwidth}
    \includegraphics[width=\textwidth]{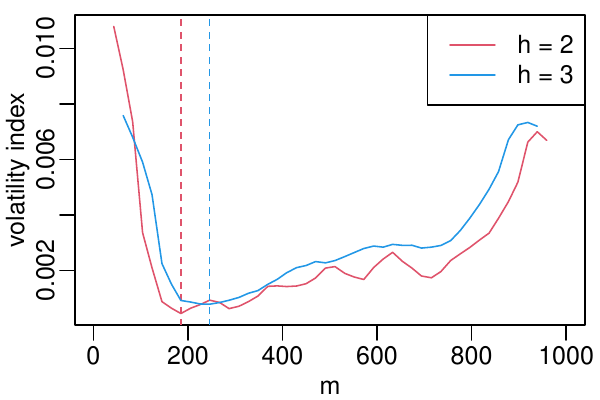}
    \subcaption{\label{fig:volatility:xicor}xicor}
  \end{subfigure}
  \begin{subfigure}[c]{0.5\textwidth}
    \includegraphics[width=\textwidth]{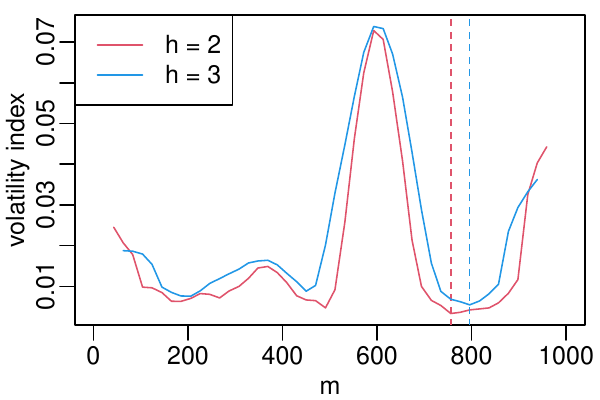}
    \subcaption{\label{fig:volatility:shorth}shorth}
  \end{subfigure}
\caption{\label{fig:volatility}Volatility index after \citet{politis:1999} as a function of $m$ for Chatterjee's correlation ({\em xicor}) and Tukey's shorth for $n=1000$.}
\end{figure}

\subsection{Choice of $m$}
According to \citet{politis:1994}, the m-out-of-n bootstrap asymptotically works for any choice of $m$ satisfying $m\to\infty$ and $m/n\to 0$ as $n\to\infty$. This, however, leaves a wide range of choices, e.g.~$m=cn^\beta$ for any $\beta\in (0,1)$, and the simulations in section \ref{sec:results} show that the convergence rate of the coverage probability depends on the choice of $m$. It might thus be desirable to choose $m$ in a data dependent way, for which different methods have been suggested in the literature \citep{politis:1999,chung:2001,goetze:2001,bickel:2008,lee:2020}.

\citet[ch.~9.3.2]{politis:1999} suggested a {\em minimum volatility method} for estimating $m$, which is based on the idea that there should be some range for $m$ where its choice has little effect on the estimated confidence interval endpoints. The optimal $m$ is that with the lowest ``volatility'' of the confidence interval, which is defined as the running standard deviation of the interval endpoints around each specific choice for $m$. Before computing the standard deviation, the interval end points should be smoothed out, too, by averaging over the endpoints computed for the neighboring choices for $m$.

The algorithm has two parameters, the window width $h_{ci}$ for smoothing the interval end points and the window width $h_\sigma$ for computing the running standard deviation. \citet{politis:1999} recommended $h_{ci}=h_\sigma=2$ or $3$, which corresponds to a range of $5$ or $7$ values. To avoid finding a minimum that is too close to $n$, as happens in Fig.~\ref{fig:volatility:shorth}, we have restricted the search range to values $m<n/2$. For efficiency reasons, \citet{politis:1999} recommended not to try every $m$, but only a grid of equidistant values. Our implementation in the function \code{estimate.m.volatility()} in the package \pkg{moonboot} therefore only tries out a maximum of 50 values for $m$.

The methods by \citet{goetze:2001} and \citet{bickel:2008} are both based on minimizing the distance between the distributions of $S_m^*$ and $S_{m'}^*$ for different choices $m$ and $m'$ of the subsampling size. As a distance measure, both articles utilized the Kolmogorov distance, i.e., the maximum distance between the cumulative distribution functions. The methods differ in the values that are tried out for $m$ and $m'$: \citet{goetze:2001} suggested to search the minimum among all values for $m$ and $m/2$, whereas \citet{bickel:2008} searched the minimum distance between $m_j$ and $m_{j+1}$ for $m_j=\lceil q^j n\rceil$, with the recommendation to set $q=0.75$. These methods are provided by \pkg{moonboot} as \code{estimate.m(..., method="goetze")} and \code{estimate.m(..., method="bickel")}, respectively.

\begin{figure}[t]
  \begin{subfigure}[c]{0.5\textwidth}
    \includegraphics[width=\textwidth]{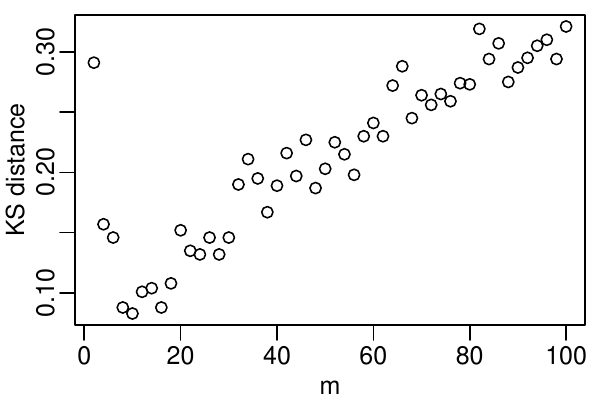}
    \subcaption{\label{fig:goetze-max:a}single sample}
  \end{subfigure}
  \begin{subfigure}[c]{0.5\textwidth}
    \includegraphics[width=\textwidth]{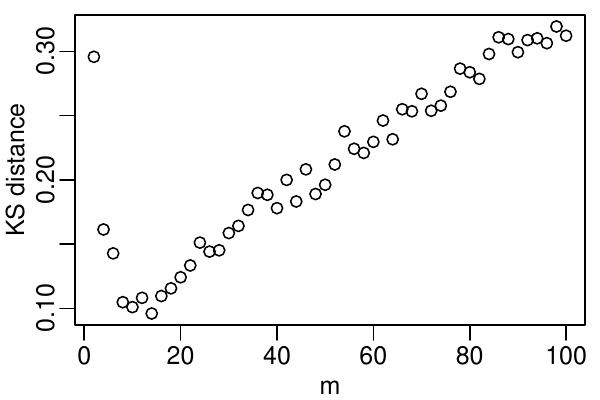}
    \subcaption{\label{fig:goetze-max:b}averaged over multiple samples}
  \end{subfigure}
\caption{\label{fig:goetze-max}Kolmogorov distance between the CDFs of $S^*_m$ and $S^*_{m/2}$ for the estimator of the maximum of a uniform distribution with $n=200$. The left results are for a single sampling per $m$, and the right results for averaging over fivefold sampling.}
\end{figure}

Trying out all combinations of $m$ and $m/2$, as suggested by \citet{goetze:2001}, has a runtime of order $O(Rn^2)$, and it would thus be preferable to use a more efficient minimum search like, e.g., a Golden Section Search \citep{press:1992}. Such algorithms, however, are devised for deterministic functions, not for random functions. As can be seen in Fig.~\ref{fig:goetze-max}, the Kolmogorov distance $d(m,m/2)$ between two randomly sampled distributions scatters considerably and this is only somewhat remedied by repeated sampling. This means that even the usual initial bracketing process by going downhill with increasing step size is unreliable when sampling from specific data, even if, on average, the function actually has a non-boundary minimum\footnote{Fig.~2 in the article by \citet{goetze:2001} shows that even this is not guaranteed, and the method can thus fail.}. We therefore sticked to the original suggestion by \citet{goetze:2001} and have implemented an exhaustive search over all even values for $m$.

Two other methods have been suggested in the literature, which we have not included in \pkg{moonboot}. \citet{chung:2001} based their method on the assumptions of root-n consistency and asymptotic normality of the estimator. Although the assumption of root-n consistency might be overcome by replacing factors $n^{1/2}$ with $\tau_n$ in the formulas, the normality assumption is essential for that method because it is based on an Edgeworth expansion. In our tests with non-normal estimators, the coverage probabilities were very low and did not even converge to the nominal value for large $n$. \citet{lee:2020} suggested to use a double bootstrap for estimating the coverage probability for different values of $m$ and to choose the $m$ where this estimate is close to the nominal value. In our experiments, however, the estimated coverage probability based on an in-sample bootstrap was either way too low or almost one and thus rarely close to the nominal value. For {\em xicor}, this behavior was not even monotone in $m$, which made the method difficult to apply. Moreover, it has a runtime complexity of $O(R^2 n^2)$, which made Monte Carlo simulations for its evaluation infeasible.

\subsection{Functions provided by \emph{moonboot}}
Like in the \pkg{boot} package, the computation of the confidence interval is split into two steps: the sampling procedure for simulating the bootstrap distribution $T_m^*$, and the computation of the confidence interval therefrom. Using the same interface makes it easy for a user to switch between \pkg{boot} and \pkg{moonboot}. Moreover, this interface is very flexible because it allows for statistics defined on complicated data structures for which only the user knows how to do the sub-indexing. The interface consists in the following two functions:

\begin{description}
\item[\code{mboot(data, statistic, m, R=1000, replace=FALSE, ...)}] \ \\
  Simulates the bootstrap distribution of the given estimator {\em statistic}, which must have been defined as a a function with two arguments: \code{statistic(data, indices)}. For multidimensional data, each row in {\em data} is assumed to be one data point. Additional arguments are passed to {\em statistic}.

\item[\code{mboot.ci(boot, conf=0.95, tau=NULL, type=c("all", "basic", "norm"))}] \ \\
  Estimates the confidence interval with Eqs.~(\ref{eq:ci-m-out-of-n}) (``basic'') or (\ref{eq:ci-norm}) (``norm''). {\em tau} must be a function that computes $\tau_n$ from its argument $n$. If it is not provided, it is estimated with \code{estimate.tau()} with the default settings of this function. According to the results in Section \ref{sec:results:tau}, this is not recommended, though, and it is preferable to provide the correct value for {\em tau}.
\end{description}

Calling these functions requires knowledge of the scaling factor $\tau_n$ and a choice for the subsample size $m$. For the cases that $\tau_n$ is not known or that $m$ shall be chosen in a data based way, two other functions are provided:
\begin{description}
\item[\parbox{\textwidth}{\raggedright\code{estimate.tau(data, statistic, R=1000, replace=FALSE, min.m=3, \\\ \ \ \ \ \ \ \ \ \ \ \ \ gamma=seq(0.2, 0.7, length.out=5), method="variance", ...)}}] \ \\[-1ex]
  Estimates the scaling factor with the method by \citet{bertail:1999}. The values for $m$ are tried out as $m_i=n^{\gamma_i}$. {\em min.m} is the minimum sample size for which the statistic makes sense and can be computed. The estimation is based on the scaling behavior of the variance (\code{method="variance"}) or of the quantile ranges (\code{method="quantile"}).

\item[\parbox{\textwidth}{\raggedright\code{estimate.m(data, statistic, tau, R=1000, replace=FALSE, min.m=3, \\\ \ \ \ \ \ \ \ \ \ \ method="bickel", params=NULL, ...)}}] \ \\[-1ex]
  Estimates $m$ with the method by \cite{bickel:2008} (\code{method="bickel"}), the method by \citet{goetze:2001} (\code{method="goetze"}), or with the volatility index (\code{method="politis"}). {\em params} is a list that can be used to pass additional parameters for the underlying method, e.g. \code{params=list(q=0.75)} for the method by \cite{bickel:2008}. {\em min.m} is the minimum sample size for which the statistic makes sense and can be computed.
\end{description}
  



\subsection{Typical usage}
Let us assume that you want estimate a confidence interval for an estimator \code{my.stat} on basis of some data \code{x}. Then you must first define a wrapper function that computes \code{my.stat} only for the data selected according to the provided indices:

\begin{example}
  boot.stat <- function(dat, indices) {
    my.stat(dat[indices])
  }
\end{example}

Without any knowledge about the asymptotic properties of \code{my.stat}, you can estimate a 95\% confidence interval with:
\begin{example}
  boot.out <- mboot(x, boot.stat, m=sqrt(NROW(x)))
  ci <- mboot.ci(boot.out, type="basic")
  print(ci)
\end{example}

Beware that this automatically estimates the asymptotic convergence rate $1/\tau_n^2$ of the variance of \code{my.stat} with the method by \citet{bertail:1999}. It is thus better, to provide this rate in the parameter \code{tau}, if you happen to know it. For a root-n consistent estimator, e.g., it is $\tau_n=\sqrt{n}$ and you can call \code{mboot.ci()} as follows\footnote{It is more efficient to just set \code{tau=sqrt}, of course, but the example uses a more complicated way for the sake of clarity.}:
\begin{example}
  ci <- mboot.ci(boot.out, tau=function(n) { n^0.5 }, type="basic")
\end{example}

\section{Examples}
\label{sec:examples}
\begin{table}[t]
  \centering\begin{tabular}{l|l|l|c}
    {\em estimator} $\hat{\theta}$ & {\em data model} & {\em true} $\theta$ & $\tau_n$ \\\hline
    \code{max(x)} & $x_i\sim\operatorname{unif}(0,1)$ & $1$ & $n$ \\
    \code{moonboot::shorth(x)} & $x_i\sim \operatorname{norm}(0,1)$ & $0$ & $n^{1/3}$ \\
    \code{XICOR::xicor(x,y)} & $x_i\sim\operatorname{unif}(-1,1)$, $y_i\sim x_i+ \mathcal{N}(0,0.5)$ & $0.3818147$ & $\sqrt{n}$ \\
    \code{mean(x)} & $x_i\sim\operatorname{power(2,0,1)}$ & $3/4$ & $\sqrt{n}$
  \end{tabular}
  \caption{\label{tbl:models}Tested estimators and data generation processes. $x=(x_1,\ldots,x_n)$ and $y=(y_1,\ldots,y_n)$ are the simulated data sets.}
\end{table}
In order to evaluate the implemented methods, we applied them to four different estimators, three of which were bootstrap inconsistent. A summary can be found in Tbl.~\ref{tbl:models}, and the detailed description follows. $x_1,\ldots,x_n$ denote the i.i.d.~sample values drawn form the given distributions.

\paragraph{Maximum of a uniform distribution (max).} The maximum likelihood estimator for the upper bound of a uniform distribution between zero and $\theta$ is $\hat{\theta}=\max\{x_1,\ldots,x_n\}$. This estimator was already given by \citet{bickel:1997} as an example for a bootstrap inconsistent estimator. It has the nice property that its probability density $g(t)$ can be readily computed as
\begin{equation}
  \label{eq:max:dist}
  g(t) = \left\{\begin{array}{ll}
  n t^{n-1}\theta^{-n} & \mbox{ for }0\leq t\leq\theta \\ 0 & \mbox{ else}
  \end{array}\right.
\end{equation}
This estimator is neither root-n consistent, nor asymptotic normal. The scaling factor $\tau_n$ can be computed from Eq.~(\ref{eq:max:dist}) as
\begin{equation}
  \label{eq:max:tau}
  \operatorname{Var}(\hat{\theta}) = \theta^2\frac{n}{(n+2)(n+1)^2} \stackrel{n\to\infty}{\sim} \frac{\theta^2}{n^2} \quad\Rightarrow\quad \tau_n = n
\end{equation}

\paragraph{Tukey's Shorth (shorth).} This is the mean of the data points in the shortest interval that contains half of the data. For symmetric distributions with a strongly unimodal density, \citet[p.~50ff]{andrews:1972} have shown that this estimator is cube root consistent, i.e., $\tau_n=n^{1/3}$. Bootstrap inconsistency of the shorth was mentioned by \citet{leger:2006}. We have implemented this estimator in \pkg{moonboot} as \code{shorth()} and applied it to normally distributed data.

\paragraph{Chatterjee's rank correlation (xicor).} This coefficient $\xi_n$ was introduced by \citet{chatterjee:2021} as an estimator for an index $\xi(X,Y)$ that is zero, if the two random variables $X$ and $Y$ are independent, and one, if $Y$ is a measurable function of $X$. For continuous $X$ and $Y$, this is an interesting example for a root-n consistent estimator \citep{lin:2022} that is bootstrap inconsistent \citep{lin:2023,dalitz:2024}. We have used the implementation \code{xicor()} from the package \CRANpkg{XICOR} \citep{xicor_r} to compute $\xi_n$, and the function given by \citet{dalitz:2024} to compute $\xi(X,Y)$ for the model $X \sim Y + \varepsilon$, where $\varepsilon$ is normally distributed with zero mean and $\sigma=0.5$. Note that we have used the original definition of $\xi_n$ by \citet{chatterjee:2021}, not the bias reduced form suggested by \citet{dalitz:2024}.

\paragraph{Mean of an unsymmetric distribution (mean).} The mean is a very well behaved statistic: it is asymptotic normal, root-n consistent and bootstrap consistent. It can thus serve as a simple test case how well the m-out-of-n bootstrap performs for bootstrap consistent estimators. We have simulated data according to the density $f(x)=3x^2$ for $x\in [0,1]$, which was already used in a study by \citet{dalitz:2017}, and for which we have implemented the random number generator \code{rpower()} in \pkg{moonboot}.

\section{Results}
\label{sec:results}
For all four estimators, we have evaluated the methods for estimating $\tau_n$ and the different choices for $m$ in the m-out-of-n bootstrap without replacement by means of Monte Carlo simulations. We have computed confidence intervals with a nominal coverage probability $P_{cov}$ of $0.95$ and estimated the actual coverage probability by repeating the computation $N=10^4$ times. This means that the accuracy (i.e. the width of a 95\% confidence interval) of the estimated $P_{cov}$ is about $\pm 0.005$. Using a larger $N$ would have made the simulation for the method by Goetze intractable. All simulations were done with $R=1000$ bootstrap repetitions.

\subsection{Estimation of $\tau_n$}
\label{sec:results:tau}
\begin{table}[t]
  \centering{\begin{tabular}{c|c|c|c||c|r||c|r}
      {\em method} & {\em estimator} & {\em true $\beta$} & $(\gamma_1,\gamma_5)$ & $n$ & {\em estimated $\beta$} & $n$ & {\em estimated $\beta$} \\\hline
      variance & mean & $0.5$ & $\mathbf{(0.2,0.5)}$ & 100 & 0.4723 & 500 & 0.4870 \\
       & &  & $(0.4,0.8)$ & & 0.5964 & & 0.5480 \\
      & max & $1.0$ & $(0.2,0.5)$ & & 0.6789 & & 0.7730 \\ 
       & &  & $\mathbf{(0.4,0.8)}$ & & 0.9595 & & 1.0043 \\ 
      & xicor & $0.5$ & $(0.2,0.5)$ & & 0.1821 & & 0.2945 \\ 
       & &  & $\mathbf{(0.4,0.8)}$ & & 0.4914 & & 0.5052 \\ \hline
      quantile & mean & $0.5$ & $\mathbf{(0.2,0.5)}$ & 100 & 0.5394 & 500 & 0.5164 \\
      range & &  & $(0.4,0.8)$ & & 0.6167 & & 0.5600 \\
      & max & $1.0$ & $(0.2,0.5)$ & & 0.8355 & & 0.8722 \\ 
       & &  & $\mathbf{(0.4,0.8)}$ & & 1.1068 & & 1.0872 \\ 
      & xicor & $0.5$ & $(0.2,0.5)$ & & 0.0332 & & 0.2128 \\ 
       & &  & $\mathbf{(0.4,0.8)}$ & & 0.4880 & & 0.5171 \\ 
  \end{tabular}}
  \caption{\label{tbl:estimatetau}Dependency of the estimation of $\tau_n=n^\beta$ on the chosen endpoints for the sequence $m_i=n^{\gamma_i}$ and two different values for $n$. The better choice is marked in bold for each estimator. The estimated values for the power $\beta$ in $\tau_n$ have been averaged over 100 estimations.}
\end{table}

\begin{figure}[b]
  \begin{subfigure}[c]{0.5\textwidth}
    \includegraphics[width=\textwidth]{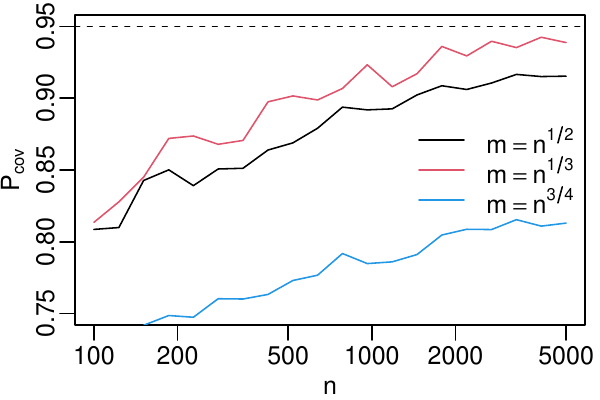}
    \subcaption{\label{fig:pcov-fixed:shorth}shorth}
  \end{subfigure}
  \begin{subfigure}[c]{0.5\textwidth}
    \includegraphics[width=\textwidth]{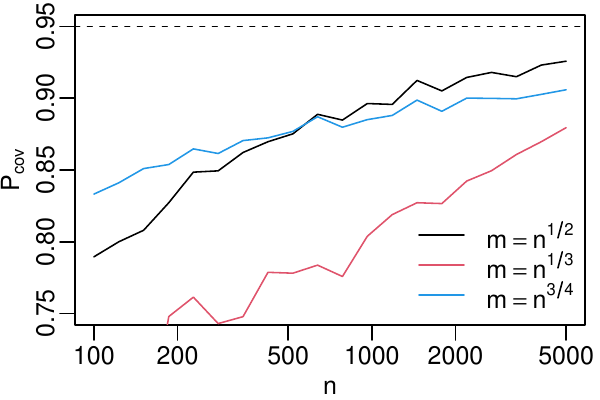}
    \subcaption{\label{fig:pcov-fixed:xicor}xicor}
  \end{subfigure}
\caption{\label{fig:pcov-fixed}Coverage probability $P_{cov}$ of the m-out-of-n bootstrap without replacement for the choices $m=n^\beta$ with $\beta\in\{\frac{1}{2},\frac{1}{3},\frac{3}{4}\}$. Note that the $n$-axis uses a logarithmic scale.}
\end{figure}

Our simulations confirmed that the choice for the extreme values of the trial values $m_i=n^{\gamma_i}$ for $i=1,\ldots,5$ indeed influence the accuracy of the estimation of $\tau_n$. For the {\em mean} estimator, e.g., we achieved the best results for $\gamma_1=0.2$ and $\gamma_5=0.5$, whereas this choice was poor for {\em xicor} and {\em max} (see Tbl.~\ref{tbl:estimatetau}). The results for estimation by means of the variance or quantile ranges are similar. We thus conclude that it does not matter which method is used. The results for {\em xicor}, however, are a warning that the estimation of $\tau_n$ with the method by \cite{bertail:1999} can be grossly inaccurate, unless the range of the $\gamma_i$ has been luckily guessed.

\subsection{Choice of $m$}
\label{sec:results:m}
To see how the coverage probability depends on the choice of $m$, we have first used the fixed formula $m=n^\beta$ with different choices for $\beta$. The results in Fig.~\ref{fig:pcov-fixed} show that the coverage probability indeed depends on the choice of $m$. Moreover, the optimal choice for $\beta$ depends on the estimator: $\beta=1/3$ was, e.g., a decent choice for {\em shorth}, but a poor choice for {\em xicor}.

\begin{figure}[t]
  \centering
  \begin{subfigure}[c]{0.49\textwidth}
    \includegraphics[width=\textwidth]{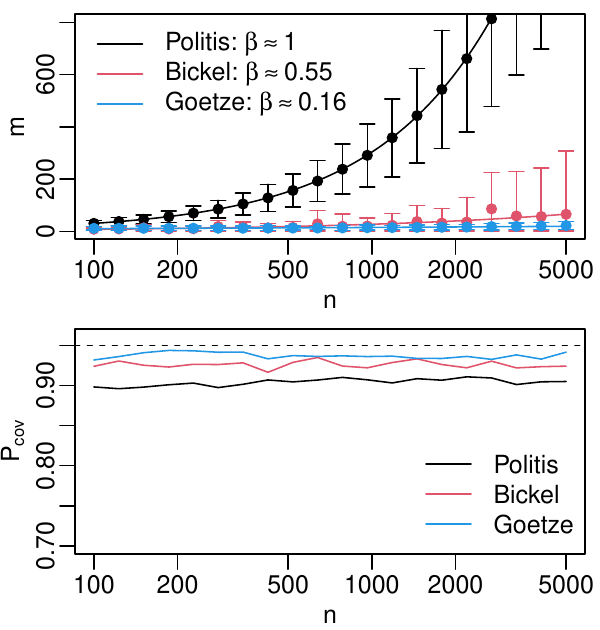}
    \subcaption{\label{fig:pcov-databased:max}max}
  \end{subfigure}
  \begin{subfigure}[c]{0.49\textwidth}
    \includegraphics[width=\textwidth]{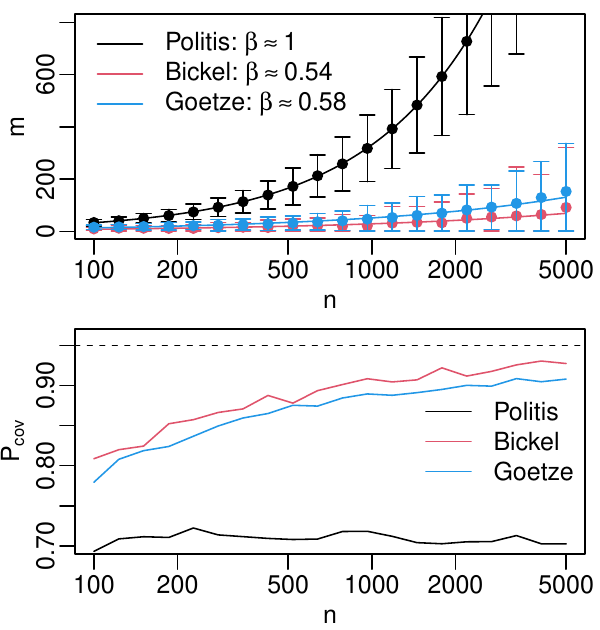}
    \subcaption{\label{fig:pcov-databased:shorth}shorth}
  \end{subfigure}

  \vspace*{2ex}
  \begin{subfigure}[c]{0.49\textwidth}
    \includegraphics[width=\textwidth]{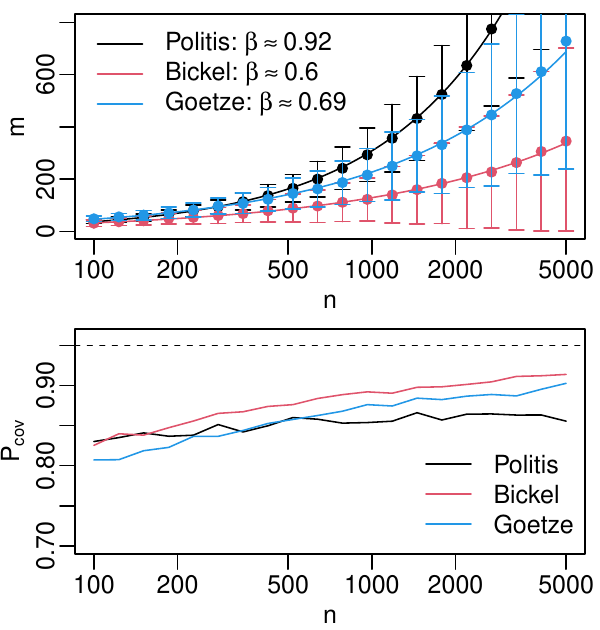}
    \subcaption{\label{fig:pcov-databased:xicor}xicor}
  \end{subfigure}
  \begin{subfigure}[c]{0.49\textwidth}
    \includegraphics[width=\textwidth]{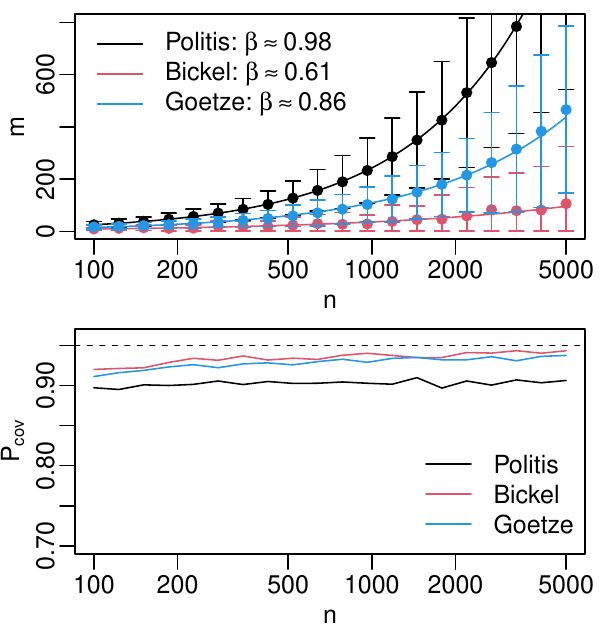}
    \subcaption{\label{fig:pcov-databased:mean}mean}
  \end{subfigure}
\caption{\label{fig:pcov-databased}Comparison of the three data based methods for choosing $m$ in the m-out-of-n bootstrap without replacement. The upper plots show the mean value $m(n)$ together with error bars $\pm\operatorname{sd}(m)$ and a least squares fit $\log(m)=\log(cn^\beta)$. The bottom plots show the resulting coverage probabilities. Note that the $n$-axis uses a logarithmic scale.}
\end{figure}

The results for the data based methods for choosing $m$ are summarized in Fig.~\ref{fig:pcov-databased}. The method by \citet{politis:1999} cannot be recommended, because the resulting coverage probability never approached the nominal value in our simulations. For the {\em shorth}, it was only about 0.7 even for high values of $n$. For the other estimators, it was somewhat greater, but nevertheless too small. This behavior can be understood from the curves showing $m(n)$ in Fig.~\ref{fig:pcov-databased}: The method by \citet{politis:1999} always chooses an $m$ proportional to $n$, which means that the condition $\lim_{n\to\infty}m/n=0$ is violated.

The methods according to \citet{goetze:2001} and \citet{bickel:2008}, on the other hand, both showed for all estimators a similar coverage probability which approached the nominal value with increasing $n$. The method by \citet{bickel:2008} was slightly better for three of the four tested estimators. It should be noted, though, that there was considerable fluctuation in the estimated values for $m$, as can be concluded from the wide error bars in Fig.~\ref{fig:pcov-databased}. As the method by \citet{goetze:2001} has a much higher runtime of order $O(Rn^2)$, the method by \citet{bickel:2008}, which has a runtime of order $O(Rn)$, is preferable from a runtime perspective, too.

\begin{figure}[t]
  \begin{subfigure}[c]{0.5\textwidth}
    \includegraphics[width=\textwidth]{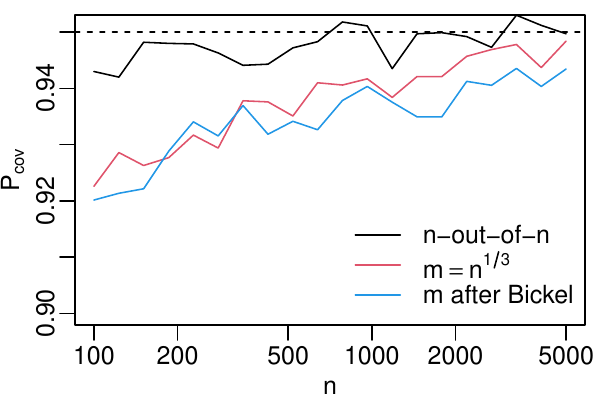}
    \subcaption{\label{fig:comparison-noon:pcov}coverage probability}
  \end{subfigure}
  \begin{subfigure}[c]{0.5\textwidth}
    \includegraphics[width=\textwidth]{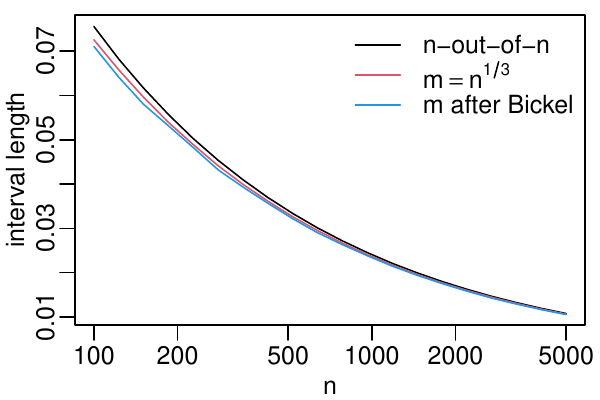}
    \subcaption{\label{fig:comparison-noon:length}average interval length}
  \end{subfigure}
\caption{\label{fig:comparison-noon}Comparison of the m-out-of-n bootstrap without replacement with the ordinary n-out-of-n (basic) bootstrap for the bootstrap consistent estimator {\em mean}. Note that the $n$-axis uses a logarithmic scale.}
\end{figure}

Out of curiosity, we have also compared the m-out-of-n bootstrap without replacement with the ordinary n-out-of-n bootstrap for the estimator {\em mean}, which is bootstrap consistent and thus allows for a comparison. For bootstrap consistent estimators with additional smoothness properties\footnote{Most notably that it is twice Fr\'echet differentiable.}, it was generally proven by \citet{bickel:1997} that estimates based on the m-out-of-n bootstrap with or without replacement are less efficient than those based on the n-out-of-n bootstrap. The mean is an example for such an estimator. As can be seen in Fig.~\ref{fig:comparison-noon}, the m-out-of-n bootstrap is competitive with the basic bootstrap only for large $n\gtrsim 2000$ and a luckily chosen $m$. For a data based choice of $m$ with the method by \citet{bickel:2008}, the coverage probability is smaller, although the difference in the interval length is small. This again demonstrates that, for bootstrap consistent estimators, the usual n-out-of-n bootstrap is preferable, especially because the best choice for $m$ is not known, in general.

\section{Violation of the assumptions}
The assumptions (\ref{eq:tau}) \& (\ref{eq:tscaled}), under which the m-out-of-n bootstrap works, are quite weak, but there are nevertheless some estimators that violate these conditions. For example, inconsistent estimators, i.e., estimators that do not converge in probability to the true parameter value, violate the assumptions because consistency is a necessary condition for the assumptions to hold, as shown in section \ref{sec:moon-bootstrap}. Using \pkg{moonboot} in such a case is an error on the side of the user, and if the user has determined the scaling factor $\tau_n$ by an analysis of the estimator, he usually will be aware of this. It might be, however, that a user relies on the fully automated estimation of $\tau_n$ provided by \pkg{moonboot} and thus is not aware of a violation of the assumptions.

We therefore have tested our package with two unbiased, yet inconsistent estimators, too. Please note that these estimators are bizarre examples and are not meant to be used in practice. The first estimator is only the very first observation as an estimator for the mean $\mu$ of the unsymmetric distribution \code{moonboot::dpower(..,2,0,1)}:
\begin{equation}
  \label{eq:mu1}
  \hat{\mu}_1=X_1
\end{equation}
The distribution of this estimator is the same for all sample sizes, and the scaling factor is thus $\tau_n\propto 1$, which fulfills condition (\ref{eq:tscaled}), but violates condition (\ref{eq:tau}). The second estimator estimates the parameter $\lambda$ of a Poisson distribution. For a Poisson distribution, both mean and variance are $\lambda$, which allows for the construction of an unbiased estimator for $\lambda$ from the unbiased estimators $\overline{X}$ for the mean and $S^2$ for the variance as follows:
\begin{equation}
  \label{eq:lambdan}
  \hat{\lambda}_n = n\overline{X} - (n-1)S^2 = \sum_{i=1}^n X_i - \sum_{i=1}^n \left( X_i - \overline{X} \right)^2
\end{equation}
As both $\overline{X}$ and $S^2$ are root-$n$ consistent estimators, the scaling factor for condition (\ref{eq:tscaled}) to hold is $\tau_n\propto n^{-1/2}$, which again is in violation of condition (\ref{eq:tau}).

As the assumptions (\ref{eq:tau}) \& (\ref{eq:tscaled}) are only sufficient, but not necessary conditions for the m-out-of-n bootstrap, we first estimated the coverage probability of the ``basic'' m-out-of-n bootstrap interval when the correct scaling factor $\tau_n$ is provided. For $\hat{\mu}_1$, the coverage probability obviously neither depends on $n$ nor on the subsampling size $m$, and it turned out to be about $0.52$ for a nominal $0.95$ interval. For $\hat{\lambda}_n$, the coverage probability decreased with increasing $n$ and was only about $0.17$ for $n=5000$ and $m=\sqrt{n}$. These results show that the m-out-of-n bootstrap indeed does not work for these inconsistent estimators.

We then estimated the scaling factors with \code{estimate.tau()} for $n=100$ and $n=500$, and, for all parameter settings, the mean estimates for $\beta$ in $\tau_n=n^\beta$ were less than $0.002$ for $\hat{\mu}_1$ and less than $-0.40$ for $\hat{\lambda}_n$. This means that, for these estimators, the violation of the assumptions can be automatically detected due to suspiciously small estimated values for $\beta$. In order to make the user aware of this issue, \code{mboot.ci()} gives a warning if $\tau_n$ decreases or if its increase is suspiciously small, i.e., if it increases at a slower rate than $n^{0.01}$.

\section{Conclusions}
The \pkg{moonboot} package provides ready-to-run implementations of the m-out-of-n bootstrap and methods for estimating its parameters. Our simulations have shown that the quality of the data based method for estimating the scaling factor $\tau_n$ is highly sensitive to a parameter for which no universal recommendations can be made. It is thus recommended to estimate $\tau_n$ by a different method, e.g. by model based Monte Carlo simulations of the variance of the estimator and estimating it by means of Eq.~(\ref{eq:vartn}). For a data based choice of $m$, the method by \citet{bickel:2008} as implemented in \code{estimate.m(..., method="bickel")} can be recommended, according to our simulations.

\section*{Acknowledgements}
We thank the anonymous reviewers for their valuable comments and for the suggestion to investigate the case when the assumptions are violated.

\bibliography{moonboot}

\address{Christoph Dalitz\\
  Niederrhein University of Applied Sciences\\
  Institute for Pattern Recognition\\
  Reinarzstr. 49\\
  Germany\\
  (ORCiD 0000-0002-7004-5584)\\
  \email{christoph.dalitz@hs-niederrhein.de}}

\address{Felix Lögler\\
  Niederrhein University of Applied Sciences\\
  Institute for Pattern Recognition\\
  Reinarzstr. 49\\
  Germany}